\documentclass{article}
\usepackage{graphicx}
\usepackage{amsmath}
\usepackage{amssymb}

\newcommand{\eqn}{equation}

\newcommand{\lb}{\left(}
\newcommand{\rb}{\right)}
\newcommand{\ph}{\hat{p}}
\newcommand{\D}{\mathcal{D}}

\newcommand{\nc}{\newcommand}

\nc{\beq}{\begin{equation}}
\nc{\eeq}{\end{equation}}
\nc{\bea}{\begin{eqnarray}}
\nc{\eea}{\end{eqnarray}}
\nc{\nn}{\nonumber}

\nc{\veps}{\varepsilon}
\nc{\eps}{\epsilon}
\nc{\as}{\alpha_s}
\nc{\cd}{\cdot}
\nc{\lag}{\cal L }
\nc{\matx}{\left|\cal {M}\right|^2}
\nc{\lqcd}{\Lambda_\textrm{QCD}}
\nc{\msbar}{\overline {\textrm{MS}}}
\nc{\really}{\stackrel{!}{=}}
\def\sla#1{\ifmmode%
\setbox0=\hbox{$#1$}%
\setbox1=\hbox to\wd0{\hss$/$\hss}\else%
\setbox0=\hbox{#1}%
\setbox1=\hbox to\wd0{\hss/\hss}\fi%
#1\hskip-\wd0\box1 }
\nc{\dsla}{\sla{\partial}}
\nc{\Dsla}{\sla{D}}

\newlength{\nseparation}
\begin{document}
\title{ 
An alternative subtraction scheme for NLO QCD calculations
}
\author{
Tania Robens\\
{\em IKTP, TU Dresden, Zellescher Weg 16, 01069 Dresden, Germany}\\
}
\maketitle
\baselineskip=11.6pt
\begin{abstract}
We discuss an alternative subtraction scheme for NLO QCD calculations, which is based on the splitting kernels of an improved parton shower.  As an example, we show results for the C parameter of the process $e^+\,e^-\,\rightarrow\,3$ jets at NLO used for the verification of this scheme.
\end{abstract}
\baselineskip=14pt

\section{Introduction}	
It is indisputable that higher order corrections are needed to correctly predict fully differential distributions for scattering processes at high precision. 
However, the implementation of NLO calculations into numerical tools exhibits a caveat stemming from the infrared divergence of real and virtual NLO contributions, which originate from different phase spaces: although in the sum of all contributions, the infinite parts exactly cancel, the behaviour of the divergence needs to be parametrized, e.g. by infinitesimal regulators. In practise, this can result in large unphysical numerical uncertainties. A way to circumvent this problem is the introduction of subtraction schemes. We here discuss a specific scheme and its properties \cite{Chung:2010fx}, using splitting kernels as well as mapping prescriptions which were already suggested in the framework of an improved parton shower \cite{Nagy:2007ty,Nagy:2008ns,Nagy:2008eq,Nagy:2014mqa}. It was further developed for processes with an arbitrary number of final states in \cite{Chung:2012rq}, with a review in \cite{Robens:2013wga}. Furthermore, the scheme has been implemented within the HelacNLO framework \cite{Bevilacqua:2013iha}.

\section{Subtraction Schemes}\label{sec:schemes}
Higher order subtraction schemes make use of factorization of the real-emission matrix element in the soft or collinear limits, leading to the decomposition
$\left|{\cal M}_{m+1}(\hat p)\right|^2 \longrightarrow \D_\ell\,\otimes\,\left| {\cal M}_{m}( p) \right|^2$
 \cite{Altarelli:1977zs,Bassetto:1984ik,Dokshitzer:1991wu}. Here and in the following, we follow the notation presented in \cite{Chung:2010fx, Chung:2012rq,Robens:2013wga}.
The subtracted contributions are then given by
\bea
\textstyle
\label{countertermfinite85}
\sigma^{\text{NLO}}&=&\underset{\text{finite}}
{\underbrace{\int_{m+1}\left[
d\sigma^R-d\sigma^A\right]}}+\underset{\text{finite}}
{\underbrace{\int_{m+1}\,d\sigma^A+\int_m\,d\sigma^V}}        
\eea
where
\begin{alignat}{53}
\label{explicitexpressionsNLO}
\textstyle
\int_m\, \left[d\sigma^B\,+\,d\sigma^V\,+\,\int_1\,d\sigma^A\right]& =\int    dPS_m
\left[\left| {\cal M}_{m} \right|^2\,+\,\left| {\cal M}_{m} \right|^2_{\text{one-loop}}\,+\,
\sum_\ell\,\mathcal{V}_\ell\,\otimes\,\left| {\cal M}_{m} \right|^2\right], \notag \\
\int_{m+1}\,\left[ d\sigma^R - d\sigma^A \right]&=\int dPS_{m+1} \left[\left| {\cal M}_{m+1} \right|^2 \,-\, \sum_\ell\,D_\ell\,\otimes\,\left| {\cal M}_{m} \right|^2\right],  
\end{alignat}
and where $\int\,d\,PS$ denotes the integration over the respective phase space, including all symmetry and flux factors.
The symbols $d\sigma^B,\,d\sigma^V,\,d\sigma^R$ stand for the Born, virtual and real-emission contributions of the calculation, while real-emission subtraction terms are summarized as $d\sigma^A$.
Since $\left|{\cal M}_{m+1}\right|^2$ and 
$\left| {\cal M}_{m} \right|^2$ live in different
phase spaces, their momenta need to be mapped via a mapping function. Furthermore, the subtraction term
$\D_\ell$ and its one-parton integrated counterpart $\mathcal{V}_\ell$ are related by 
$\mathcal{V}_\ell\,=\, \int\,d\xi_p \,\D_{\ell},$
where $d\xi_p$ is an unresolved one-parton integration measure. 
In the scheme discussed here,
we apply a momentum mapping which leads to an overall scaling behaviour $\sim\,N^2$ for a process with $N$ partons in the final state.

\section{Scheme setup}\label{sec:ns-sub}
We denote
four-momenta in the Born-type kinematics by unhatted quantities $p_i$, while the real emission phase space momenta are denoted by hatted quantities $\ph_i$;
initial state momenta are labelled $p_a$ and $p_b$, where $Q\,=\,p_a+p_b$ and with $Q^2$ being the squared centre-of-mass energy, with equivalent relations in the real emission phase space;
generally, $\ph_\ell$ labels the emitter, $\ph_j$ the emitted parton and $\ph_k$ the spectator.

The real emission matrix element $\mid {\cal M}_\ell(\{\hat p, \hat f\}_{m+1})\rangle$ is related to the Born one $\mid{\cal M}(\{ p,  f\}_{m})\rangle$ via \cite{Nagy:2007ty}
\beq
\label{QCDFactorizationm1tVm}
\mid {\cal M}_\ell(\{\hat p, \hat f\}_{m+1})\rangle\,=\,t^\dagger_\ell(f_\ell \to \hat f_\ell + 
\hat f_{j})\,V^\dagger_\ell(\{\hat p, \hat f\}_{m+1})\,\mid {\cal M}(\{ p,  f\}_{m})\rangle,
\eeq

In our scheme, soft/ collinear divergences from interference terms are treated using dipole partitioning functions $A_{\ell k}$ \cite{Nagy:2008eq}, which have been explicitely discussed in \cite{Chung:2010fx,Chung:2012rq,Robens:2013wga}.
All (integrated) subtraction terms are specified in the same reference.

The improved scaling behaviour of our scheme mainly results from the specific mapping between the real emission and Born-type kinematic phase spaces for final state emitters. For final state mappings, we use the whole remainder of the event as a spectator in terms of momentum redistributions:
\begin{eqnarray}
\label{eq:fin_map}
p_\ell\,=\,\frac{1}{\lambda_\ell}\,(\hat p_\ell + \hat p_j)-\frac{1 - \lambda_\ell + y_\ell}{2\, \lambda_\ell\, a_\ell}\, Q,\;\;
p_n^\mu\,=\,\Lambda (K,\hat{K})^\mu{}_\nu \,\hat p^{\nu}_{n} ,\quad n\notin\{\ell,j=m+1\},
\end{eqnarray}
with
${\textstyle \Lambda(K,\hat K)^{\mu}_{\;\;\nu} \,=\,g^{\mu}_{\;\;\nu}\,-\,\frac{2\,( K+\hat K)^{\mu}\,(K+\hat K)_{\nu}}{(K+\hat K)^{2}}\,
+\,\frac{2\,{K}^{\mu}\,\hat{K}_{\nu}}{\hat K^{2}}\,},$
where
${\textstyle
y_\ell = \frac{P_\ell^2}{2\, P_\ell\cdot Q - P_\ell^2}.}$
where we introduced
${\textstyle \lambda_\ell\lb y_\ell,a_\ell \rb\,=\,\sqrt{\left(1+y_\ell\right)^2-4\,a_\ell\,y_\ell},}\;$ ${\textstyle K\,=\, Q - p_\ell,}\;$ ${\hat{K}\,=\, Q - P_\ell,\,a_\ell\lb P_\ell,Q\rb\,=\,\frac{Q^2}{2\, P_\ell\,\cdot\,Q-P_\ell^2}}$, with $P_\ell\,=\,\ph_\ell+\ph_j$. 
It is the {\sl global} mapping for all remaining particles in Eqn. (\ref{eq:fin_map}) that is responsible for the reduced number of Born-type matrix reevaluations. 
For the real emission subtraction terms, we then obtain the total contribution
\begin{\eqn}\label{eq:master_sub}
d\sigma^{A}_{ab}(\ph_a,\ph_b)\,=\,d\sigma^{A,a}_{ab}(\ph_a,\ph_b)+d\sigma^{A,b}_{ab}(\ph_a,\ph_b)
+\sum_{\ell\,\neq\,a,\,b} d\sigma^{A,\ell}_{ab}(\ph_a,\ph_b),
\end{\eqn}
with the sum over all possible final state emitters.

In the setup of the scheme, the finite remainders of some subtraction terms are currently evaluated numerically. This poses no impediment for the implementation of our scheme. We have approximated all remainders for numerical integrals by approximation functions, cf. \cite{Bach:2013sha} for a first preliminary discussion.

\section{Results}\label{sec:status}
 We here show the results for the C parameter in the process $e^+\,e^-\,\rightarrow\,\text{3 jets}$ \cite{Chung:2012rq}.
For this, 
the real emission processes are given by
\begin{eqnarray}
&&e^+\,e^-\,\rightarrow\,q\,\bar{q}\,q\,\bar{q},\;e^+\,e^-\,\rightarrow\,q\,\bar{q}\,g\,g.
\end{eqnarray}
These contributions call for $(8+10)$ matrix element reevaluations per phase space point in the Catani-Seymour \cite{Catani:1996vz} and $(4+5)$ reevaluations in our scheme, respectively.
We display our results in terms of the C distribution\cite{Ellis:1980wv} 
\begin{\eqn}\label{eq:C}
{\textstyle C^{(n)}\,=\,3\,\left\{ 1-\sum_{i,j\,=\,1,\,i<j}^n\,\frac{s_{ij}^2}{(2\,p_i\cdot\,Q)\,(2\,p_j\cdot\,Q)}  \right\},\,{(s_{ij}\,=\,2\,p_i \cdot p_j)}}.
\end{\eqn}

\begin{figure}
\begin{minipage}{0.49\textwidth}
\begin{center}
\includegraphics[width=\textwidth]{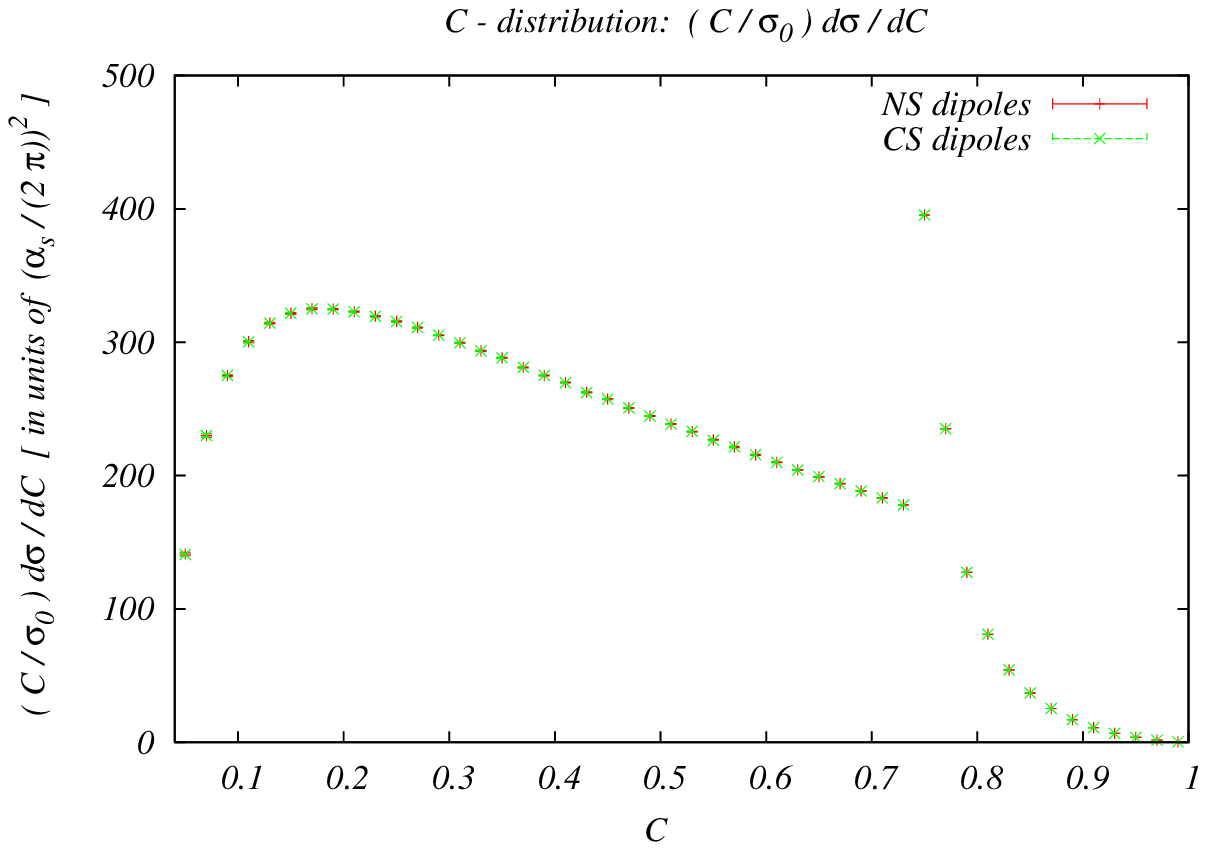}
\end{center}
\end{minipage}
\begin{minipage}{0.49\textwidth}
\begin{center}
\includegraphics[width=\textwidth]{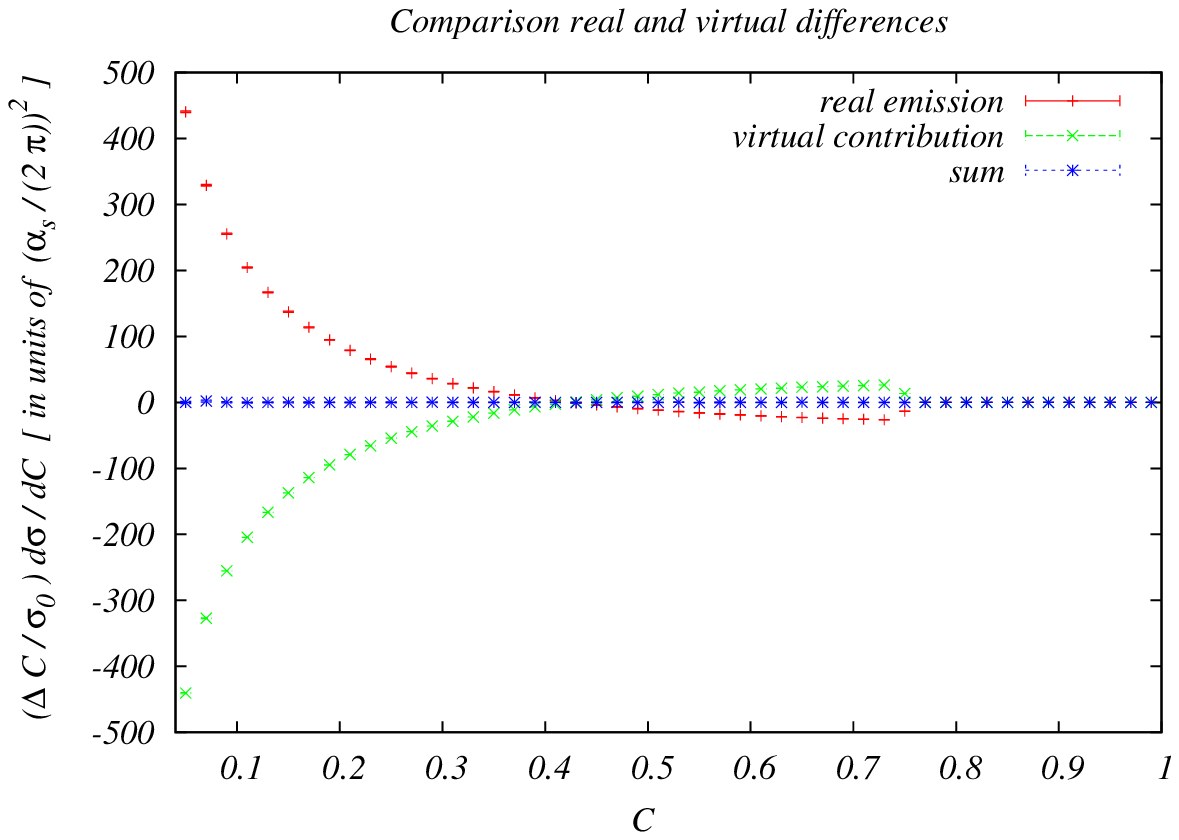}
\end{center}
\end{minipage}
\caption{\label{fig:tot_diff} {\sl Left:} Total result for differential distribution $\frac{C}{\sigma_0}\,\frac{d\sigma^\text{NLO}}{dC}$ using both our dipoles (red, "NS")  and Catani-Seymour dipoles (green, "CS"). The standard literature result obtained using the CS scheme is completely reproduced with the NS dipoles. {\sl Right:} {\sl Differences} $\Delta_\text{CS-NS}$ for real emission (red, upper) and virtual (green, lower) contributions, showing that especially for low $C$ values the contributions in the two schemes significantly differ. Adding up $\Delta^\text{real}+\Delta^\text{virt}$ renders 0 as expected.}
\end{figure}
Figure \ref{fig:tot_diff} shows that we reproduce the literature result \cite{mike}, as well as agreement between implementations of both schemes. We want to point out that this is indeed a non-trivial statement, as the {\sl differences} between the two schemes for both subtracted real emission as well as virtual contributions are sizeable.

\section{Summary}

We  here reported on an alternative NLO subtraction scheme for QCD calculations, which uses the splitting functions of an improved parton shower as subtraction kernels. We have briefly discussed the setup, and especially the features leading to an improved scaling behaviour of our scheme. Results for the process $e^+\,e^-\,\rightarrow\,3$ jets have been presented. Summarizing, we regard the scheme discussed here as a viable alternative to standard schemes.
\section{Acknowledgements}
TR wants to thank the organizers for an excellent workshop at ECT$^*$.

\bibliographystyle{hunsrt}
\bibliography{NLO_subtraction}

\end{document}